\documentstyle[epsfig,12pt]{article}

\newcommand{\bce}{\begin{center}} 
\newcommand{\ece}{\end{center}}
\newcommand{\beq}{\begin{equation}}
\newcommand{\eeq}{\end{equation}}
\newcommand{\bea}{\vspace{0.25cm}\begin{eqnarray}}
\newcommand{\eea}{\end{eqnarray}}

\newcommand{\ba}{\begin{array}}
\newcommand{\ea}{\end{array}}

\newcommand{\ket}[1]{| {#1} \rangle}
\newcommand{\bra}[1]{\langle {#1} |}

\newcommand{\doublespace}{
    \renewcommand{\baselinestretch}{1.6}\large\normalsize}

\def\lsim{\mathrel{\rlap{\lower4pt\hbox{\hskip1pt$\sim$}}
    \raise1pt\hbox{$<$}}}         
\def\gsim{\mathrel{\rlap{\lower4pt\hbox{\hskip1pt$\sim$}}
    \raise1pt\hbox{$>$}}}         

\def\beq{\begin{equation}}
\def\endeq{\end{equation}}
\def\arr{\begin{eqnarray}}
\def\endarr{\end{eqnarray}}

\makeindex

  \advance\oddsidemargin  by -1in
  \advance\evensidemargin by -1in
  \advance\topmargin      by -0.5in
\begin{document}

\vspace{2.0cm}

\begin{flushright}
\end{flushright}

\vspace{1.0cm}

\begin{center}
{\Large \bf 
Recovering partial conservation of  axial current\\
 in diffractive neutrino scattering}\\

\vspace{1.0cm} 
{\large \bf  V.A. Novikov$^{1,2,3}$ and  
V.R. Zoller$^1$}\\
\vspace{0.5cm}
$^1${\em Institute for  Theoretical and Experimental Physics,
Moscow 117218, Russia}\\
$^2${\em  MPTI, Moscow, Russia}\\
$^3${\em NSU, Novosibirsk, Russia}
\vspace{2.0cm}

{ \bf Abstract }\\
\end{center}


A  model of diffractive neutrino scattering is formulated in terms of
 the chiral hadronic current which is conserved in the  limit of 
vanishing pion mass.  
This current has the correct singularity structure and, naturally, 
 does not lead to contradictions with a partial conservation of 
the axial current (PCAC). In that respect we differ from earlier work
 in the literature, where a breakdown of PCAC 
had been reported. We show that such a breakdown of PCAC is an 
artifact of the hadronic current non-conservation in the model 
developed there.

\doublespace

\vskip 0.5cm \vfill $\begin{array}{ll}
\mbox{{\it email address:}} & \mbox{novikov@itep.ru} \\
\mbox{{\it email address:}} & \mbox{zoller@itep.ru} \\
\end{array}$

\pagebreak


\section{Introduction}

This communication is motivated by the publication \cite{Kop} entitled
``Breakdown of partial conservation of axial 
current  in diffractive neutrino scattering''. 
The analysis \cite{Kop}
is based on a specific model of diffractive neutrino scattering 
suggested 
earlier in Ref.\cite{STOD}. Within this model interactions 
of high-energy neutrino with the nucleon or nuclear target
\beq
\nu+N\to l +X
\label{eq:nutol}
\eeq
in the axial channel are mediated by pions 
and $a_1(J^{PC}=1^{++})$ mesons. 
 Corresponding matrix  elements of the axial hadronic current, $A_{\mu}$, 
are expanded over $\pi$ and $a_1$ components. 
For the specific final state $\ket{X}=\ket{\pi N}$  
this expansion, with certain reservations, leads to the requirement 
{\footnote{In addition to the longitudinal $a_1$, in Ref.~\cite{Marage},
 contributions of the 
 $\rho-\pi$ state
and  higher axial excitations where also  considered.}
\beq
\sigma(\pi N\to \pi N)=\sigma(\pi N\to a_1 N).
\label{eq:PSEUDO}
\eeq 
The latter equality is considered in  \cite{Kop} as an 
indispensable property of a partial conservation of the axial current 
(PCAC). 
In \cite{Kop} it was found   
that Eq.(\ref{eq:PSEUDO}) can not  be reconciled with experimental
data and the breakdown of PCAC was claimed. 

Below we show that Eq.(\ref{eq:PSEUDO}) does not follow from PCAC
and can not be a basis for  radical  questioning of 
PCAC. 

\section{The $\pi-a_1$-model and $a_1$-dominance}

Below in Sects.2 and 3 we briefly sketch the derivation of 
Eq.(\ref{eq:PSEUDO}). For more details see \cite{STOD,Marage}.

Within the $\pi-a_1$-model developed in \cite{STOD} and exploited in
\cite{Kop,Marage}
the matrix element of the hadronic axial current
\beq 
A_{\mu}=\bra{X}a_{\mu}(0)\ket{N}
\label{eq:Amua}
\eeq
 entering  the amplitude
\beq
T(\nu N\to l X)={G_F\over \sqrt{2}}L_{\mu}(V_{\mu}+A_{\mu}),
\label{eq:lVA}
\eeq
of the process (\ref{eq:nutol}) is saturated by the two lowest
hadronic states, $\pi$ and $a_1$ mesons,
\beq
A_{\mu}=f_{\pi}{q_{\mu}\over {q^2-\mu^2}}T(\pi N\to X)
+f_a{{g_{\mu\nu}-q_{\mu}q_{\nu}/M^2}\over q^2-M^2}T_{\nu}(a_1 N\to X).
\label{eq:ASTOD}
\eeq
Hereafter, $\mu$ stands for the pion mass and $M$ - for the $a_1$.

One comment on Eq.(\ref{eq:ASTOD}) is in order.
This equation
provides 
the off-mass-shell extrapolation of physical amplitudes $\pi N\to X$ and 
$a_1 N\to X$. Far from the $\pi$-,$a_1$-pole 
 the representation (\ref{eq:ASTOD})  becomes
rather uncertain at least for the $a_1$-exchange which is always very far
the mass shell in
the reaction (\ref{eq:nutol}).
To minimize uncertainties in  Sect.~5 we make  use of  the symmetry property 
of the problem, The latter turns out to be crucial
for (in)validity of  Eq.(\ref{eq:PSEUDO}).

The  leptonic current
\beq
L_{\mu}=\bar{u}(k^{\prime})\gamma_{\mu}(1+\gamma_{5})u(k).
\label{eq:lmu}
\eeq
is purely transversal (we neglected the lepton mass, 
$m_l=0$, and introduced $q=k-k^{\prime}$), 
\beq
q_{\mu}L_{\mu}=0.
\label{eq:LTRANS} 
\eeq
From Eq.(\ref{eq:LTRANS}) it follows that
  the pion pole  in Eq.~(\ref{eq:ASTOD}) 
 does not contribute to the $\nu p$-scattering cross-section.
 Then,
 the longitudinal component of the
differential cross section of 
the process (\ref{eq:nutol}) within the $\pi-a_1$-model of 
Ref.~\cite{STOD} (see also \cite{Marage}) is dominated by 
the $a_1$ contribution,
\beq
{d^2\sigma(\nu p\to l X)\over dQ^2d\nu}\propto 
{f^2_aQ^2\over M^4}\sigma^L(a_1 p\to X;\,Q^2),
\label{eq:A1TOX}
\eeq
where we denoted $Q^2=-q^2$.
{\footnote{We are interested in the limit $Q^2\to 0$ where 
$\sigma^L(Q^2)$ is singular. Recall that in the axial channel 
$
\sigma^L=|\epsilon^L_{\mu}A_{\mu}|^2/\sqrt{Q^2+\nu^2}
$
and the longitudinal polarization vector is defined as
$
\epsilon^L=(\sqrt{\nu^2+Q^2},0,0,\nu)/\sqrt{Q^2}
$
with $q=(\nu,0,0,\sqrt{\nu^2+Q^2})$.}}

Adler's observation is that the above differential  cross section 
(\ref{eq:A1TOX}) at $Q^2\to 0$
 is expressible also in terms of the on-shell pion-nucleon 
cross section $\sigma(\pi N\to X)$  \cite{Adler}.

\section{Pions  and   Adler's theorem }

In Ref.~\cite{Adler} it was noticed that at $Q^2=0$ 
\beq
L_{\mu}\propto q_{\mu}.
\label{eq:QUMU}
\eeq
Consequently,
\beq
T(\nu p\to l X)\propto q_{\mu}A_{\mu}.
\label{eq:MADLER}
\eeq
The constraint of PCAC implies \cite{Nambu}
\beq
 q_{\mu}a_{\mu}=f_{\pi}\mu^2\varphi,
\label{eq:PCAC}
\eeq
where $f_{\pi}$ is the pion decay constant,  $\varphi$ is the pion 
field operator and $a_{\mu}$ is the axial current operator (see Eq.(\ref{eq:Amua})).
Therefore,
\beq
 |q_{\mu}A_{\mu}|^2={f^2_{\pi}\over \sqrt{\nu^2+Q^2}}\sigma(\pi N\to X)
\label{eq:QUMU2}
\eeq
and at $Q^2\to 0$
\beq
{d^2\sigma(\nu p\to l X)\over dQ^2d\nu}\propto 
{f^2_{\pi}}\sigma(\pi N\to X)
\label{eq:PITOX}
\eeq

In Ref.\cite{STOD} (see also \cite{Marage}) from 
comparison of Eqs.(\ref{eq:PITOX}) and (\ref{eq:A1TOX}) 
supplemented with Weinberg sum rules \cite{Wein} and 
 certain assumptions on the off-shell properties of  hadronic 
cross sections   Eq.(\ref{eq:PSEUDO}) was obtained.

\section{The $\pi-a_1$-model - the model with  built-in current non-conservation}
In \cite{Kop} it was noticed that 
 the cross sections $\sigma(\pi N\to \pi N)$ 
and $\sigma(\pi N\to a_1 N)$, 
where $N$ represents the target nucleon/nucleus, 
have different dependence on the collision
 energy as well as  very different dependence on the nuclear opacity. 
The principal conclusion of Ref.\cite{Kop} is that the PCAC hypothesis
 is in conflict with well established properties of high-energy 
 hadronic amplitudes.
 
However, it is quite clear that the basic  expansion (\ref{eq:ASTOD}) 
has at least one serious flaw.
The current (\ref{eq:ASTOD}) is not conserved. It is not conserved even
``partially''. Consequences  are obvious.
The requirement of PCAC (\ref{eq:PCAC}) 
supplemented with the equation of motion of the pseudoscalar 
field $\varphi$ applied to the 
matrix element (\ref{eq:Amua}) implies
\beq
q_{\mu}A_{\mu}=f_{\pi}{\mu^2\over{q^2-\mu^2}}T(\pi N\to X).
\label{eq:DERIV}
\eeq
and for $A_{\mu}$ defined  by Eq.(\ref{eq:ASTOD}) results in
\beq
f_{\pi}T(\pi N\to X)=f_a M^{-2}q_{\nu}T_{\nu}(a_1 N\to X).
\label{eq:STODEQ}
\eeq
Eq.(\ref{eq:STODEQ}) like its counterpart (\ref{eq:PSEUDO}) can hardly be
reconciled   with the experimental data.  

\section{Introducing the chiral hadronic current $A^{\chi}_{\mu}$}

To meet the requirement of chiral symmetry the matrix element of
the axial hadronic current in the 
basis of
$\pi,\,a_1$-states should be constructed as follows 
\beq
A^{\chi}_{\mu}=g_A{M^2\over {q^2-M^2}}\left[g_{\mu\nu}
-{q_{\mu}q_{\nu}\over q^2-\mu^2}\right]{\cal T}_{\nu}(q^2,...)
\label{eq:ACHIRAL}
\eeq
Below we keep in  ${\cal T}_{\nu}(q^2,...)$  
only one argument. The dependence on additional variables arises in 
specific problems for particular final states. 

Two poles in (\ref{eq:ACHIRAL}) correspond to both the  pion and the 
$a_1$-meson. At $q^2\approx \mu^2$
\beq
A^{\chi}_{\mu}=g_A\left[g_{\mu\nu}
-{q_{\mu}q_{\nu}\over q^2-\mu^2}\right]{\cal T}_{\nu}(q^2)
\label{eq:ACHIRALPI}
\eeq
and for $q^2\approx M^2$
\beq
A^{\chi}_{\mu}=g_A{M^2\over {q^2-M^2}}\left[g_{\mu\nu}
-{q_{\mu}q_{\nu}\over M^2}\right]{\cal T}_{\nu}(q^2),
\label{eq:ACHIRALA1}
\eeq
as it should be. In the limit $\mu\to 0$
the current (\ref{eq:ACHIRAL}) is conserved,
\beq
q_{\mu}A^{\chi}_{\mu}=0.
\label{CHICONS}
\eeq

The current conservation is, thus, a purely kinematical effect.
The  dynamics is concentrated in
the invariant amplitude  ${\cal T}_{\nu}(q^2)$,
which  is  controlled by the QCD. The latter implies that 
in the particular two-channel model one and the same  function
  ${\cal T}_{\nu}(q^2)$ describes neutrino scattering in a wide range 
of virtualities $q^2$ including both $\pi$ and $a_1$ poles, where
${\cal T}_{\nu}(q^2)$
 satisfies the following on-shell conditions
\beq
g_A q_{\nu}{\cal T}_{\nu}(\mu^2)=f_{\pi}T(\pi N\to X),
\label{eq:PROPER1}
\eeq
\beq
g_A M^2 {\cal T}_{\nu}(M^2)=f_a T_{\nu}(a_1 N\to X).
\label{eq:PROPER2}
\eeq

The product of leptonic and hadronic currents - 
recall that $q_{\mu}L_{\mu}=0$, - is as follows
\beq
T(\nu N\to l X)\sim L_{\mu}A^{\chi}_{\mu} = 
g_A L_{\mu}{\cal T}_{\mu}(q^2).
\label{eq:AMPL}
\eeq
Here, ${\cal T}_{\mu}(q^2)$ is specified by 
Eqs.(\ref{eq:PROPER1},\ref{eq:PROPER2}).
 Comparison of Eq.(\ref{eq:AMPL}) at $q^2\to 0$ with Adler's amplitude 
dictated by PCAC 
 leads simply to the identity (\ref{eq:PROPER1}) and does not yield
any new relation with $T(a_1 N\to X)$
 because now the off-shell extrapolations of  amplitudes 
 $T(\pi N\to X)$ and $T(a_1 N\to X)$
are interrelated by the current conservation condition. Indeed,
the chiral current $A^{\chi}_{\mu}$ can be represented
as a superposition of $\pi$ and $a_1$ poles. In the limit of $\mu^2\to 0$
\bea
A_{\mu}^{\chi}=g_A {q_{\mu}q_{\nu}\over {q^2-\mu^2}}{\cal T}_{\nu}(q^2)+\nonumber\\
+g_A {M^2\over {q^2-M^2}}\left[g_{\mu\nu}
-{q_{\mu}q_{\nu}\over M^2}\right]{\cal T}_{\nu}(q^2).
\label{eq:EXPAND}
\eea
Eq.(\ref{eq:ASTOD}) follows from Eq.(\ref{eq:EXPAND})  if  
${\cal T}_{\nu}(q^2)$ 
is substituted with two  different on-shell amplitudes.
 Evidently, this operation breaks down PCAC.

\section{Summary and Conclusions}

The breakdown of PCAC claimed in \cite{Kop} has been derived 
from the $\pi-a_1$-model of Ref.\cite{STOD}.
An important ingredient of the model is the matrix element of the 
 axial current (\ref{eq:ASTOD})
which is saturated by the lowest axial hadronic states, 
$\pi$ and $a_1$. In the kinematical domain of the reaction (\ref{eq:nutol})
the exchanged $a_1$-meson is always very far from the mass shell.
Postulated in \cite{STOD,Marage} the off-shell extrapolation of
(\ref{eq:ASTOD})  has serious flaw, the current (\ref{eq:ASTOD}) 
is not conserved. 
The model \cite{STOD} simply does not respect the 
chiral symmetry. No wonder, the PCAC in the model is  badly broken.

We introduced the  axial hadronic
current $A^{\chi}_{\mu}$ (see Eq.(\ref{eq:ACHIRAL})). 
This current  is conserved in the chiral limit. Also it has a
correct  $\pi-a_1$-pole structure. Naturally, this current does not
lead to any troubles with PCAC.

{\bf Acknowledgments.} 

Thanks are due  to N.N. Nikolaev for careful
 reading the manuscript.
The work was supported in part by
 the RFBR grants 11-02-00441 and 12-02-00193.


\begin{thebibliography}{299}


\bibitem{Kop}
B.Z. Kopeliovich, I.K. Potashnikova, Ivan Schmidt and M. Siddikov, 
Phys. Rev. C{\bf 84}, 024608 (2011). 


\bibitem{STOD} C.A. Piketty and L. Stodolsky,  {Nucl. Phys.} B\textbf {15}, 
571 (1970).
 
\bibitem{Marage} B.Z. Kopeliovich and P. Marage,  {Int. J. Mod.  Phys.} 
A\textbf {8}, 1513 (1993);
A.A.~Belkov and B.Z.~Kopeliovich, Sov.J.Nucl.Phys.
 \textbf{46}, 499 (1987).

\bibitem{Adler} S. Adler, {Phys. Rev.} \textbf{135}, B963 (1964).

\bibitem{Nambu} Y. Nambu, {Phys. Rev. Lett.} \textbf{4}, 380 (1960);
M. Gell-Mann and M. Levy, Nuovo Cimento, \textbf{17}, 705 (1960).

\bibitem{Wein} S. Weinberg, {Phys. Rev. Lett.} \textbf{18}, 507 (1967).


\end{thebibliography}
\end{document}